\documentclass[12pt]{article}

\begin{document}

\begin{center}
{\LARGE More about Dynamical Reduction and \\ the Enumeration
Principle \par}
\lineskip .75em
{\Large \emph{Angelo Bassi and GianCarlo Ghirardi} \par} \vskip .75em
\end{center}
\begin{center}
\Large ABSTRACT
\lineskip .75em
\end{center}

In view of the arguments put forward by Clifton and Monton [1999]
in a recent preprint, we reconsider the alleged conflict of
dynamical reduction models with the enumeration principle. We
prove that our original analysis of such a problem is correct,
that the GRW model does not meet any difficulty and that the
reasoning of the above authors is inappropriate since it does not
take into account the correct interpretation of the dynamical
reduction theories.

\vskip .5em
\vskip .5em
\vskip .5em
\noindent 1 \textit{Introduction}
\vskip .5em
\noindent 2 \textit{A Brief Account of the Enumeration Anomaly}
\vskip .5em
\noindent 3 \textit{The Interpretation of the GRW Theory}
\vskip .5em
\noindent 4 \textit{A Simplified Realistic Example of the GRW Dynamics for
a Macroscopic
\\ \,\,\,\,\,\,\,\, Object}
\vskip .5em
\noindent 5 \textit{A Critical Analysis of the Request to Operationalise
the Process of
Counting Marbles}
\newpage

\section{Introduction}

In a recent paper (Lewis [1997]) the compatibility of dynamical
models of spontaneous reduction (in particular of the so called GRW model
(Ghirardi, Rimini, Weber [1986])) with the enumeration principle has been
questioned. We
have replied to the arguments of Lewis [1997] on this journal (Ghirardi,
Bassi [1999]).
Subsequently, we have received a preprint by R. Clifton and B. Monton [1999] in
which the whole matter is reconsidered. Their conclusions can be
summarized in the following terms: the authors agree with us that Lewis'
criticisms to
GRW are not cogent but they consider our argument  as wrong and
feel the necessity of presenting the \textit{correct way} to derive the desired
conclusion. Such a way requires a quite elaborated (even though in our
opinion not appropriate and superfluous) argument proving that the `counting
anomaly' cannot, in principle, be put into evidence. In this paper we
reconsider the whole matter pointing out that  both  Lewis [1997] and
Clifton and
Monton [1999] have not taken the correct attitude about the interpretation
of the GRW
theory and that the central proof of Clifton and Monton is quite inappropriate.
We strengthen our argument by resorting to explicit examples to make clear
to the
readers some specific dynamical aspects of reduction theories which seem to
have been
misinterpreted, and which must be taken into account when dealing with systems
containing an enormous number of macroscopic subsystems.

\section{A Brief Account of the Enumeration Anomaly}

The essential point which has given rise to various critical remarks
(Shimony [1991];
Albert and Loewer [1996]) about dynamical reduction theories and is the
very starting
point of the arguments of Lewis [1997] and Clifton and Monton [1999] is the
so called
`tails problem'. It consists in the fact that the collapse processes
characterizing the GRW theory (as well as its more refined versions such as
CSL (Pearle [1989]; Ghirardi, Pearle, Rimini [1990]) unavoidably lead to
wavefunctions
which have a non compact support in configuration space, thus giving rise
to problems with
the attribution of determined locations to physical systems.

The alleged puzzling implications of this fact for the problem we are
interested in can be exemplified in a very simple way. One considers a
macroscopic marble (which, for simplicity, we take to have a radius of about
one centimeter and normal density) in an initial state at time $t=0$ denoted
as
\begin{equation}
|\Psi (0)\rangle =\frac{1}{\sqrt{2}}[\left| in\right\rangle +\left|
out\right\rangle ],  \label{1}
\end{equation}
the states $\left| in\right\rangle $ ($\left| out\right\rangle $) being
eigenstates of the marble being well inside (outside) an extremely large
box, and one recalls, as just mentioned, that dynamical reduction theories
lead to wavefunctions which, in the case of macroscopic objects, are
certainly very well localized but always possess `tails' going off to
infinity. To give a sinthetic but significative description of the state of
affairs one must recall that GRW's dynamics implies that a macroscopic object
cannot remain (Bell [1987]) for \textit{more than a split second} in a state
like (1), but it will always be in one of the following two states:
\begin{equation}
|\Psi _{in}(t)\rangle =\alpha \left| in\right\rangle +\beta \left|
out\right\rangle ,  \label{2a}
\end{equation}
or
\begin{equation}
|\Psi _{out}(t)\rangle =\beta \left| in\right\rangle +\alpha \left|
out\right\rangle,
\label{2b}
\end{equation}
with $\left| \alpha \right| ^{2}+\left| \beta \right| ^{2}=1,$ $\left|
\alpha \right| ^{2}>>>\left| \beta \right| ^{2}.$ Lewis [1997] and Clifton
and Monton
[1999] agree that one can interpret state (\ref{2a}) as one in which the
marble \emph{is}
inside the box, and state (\ref{2b}) as one in which it\emph{\ is} outside
the box, even
though in the two papers slightly different reasons are presented to
justify the above
claims. Lewis [1997] resorts to the consideration of the scalar product to
decide when
two states are `near to each other' and consequently can be considered as
describing
similar properties:

\begin{quotation}
Those who defend the GRW theory point out that (2) is very close to the
state $|in\rangle $ in which the marble is determinately in the box, and (3)
is very close to the eigenstate $|out\rangle $ in which the marble is
determinately outside the box. The scalar product provides a convenient
measure of the proximity of the two states ...

Intuitively, it seems that
if two states are sufficiently close, they will be macroscopically
equivalent, at least for all practical purposes (Lewis [1997] p.316).
\end{quotation}
On the other hand Clifton and Monton [1999] adopt the criterion PosR
proposed by
Albert and Lower [1996] to decide whether a particle is inside or
outside a given box:

\begin{quotation}
`Particle $x$ is in region $R$' if and only if the proportion of the total
squared amplitude of $x$'s wave function which is associated with points in $%
R$ is greater than or equal to $1-p $ (Clifton and Monton [1999] p.4),
\end{quotation}
$p$ being an appropriately chosen positive number smaller than $0.5.$

One then takes into account a system of a large number $n$ of
non-interacting marbles, each of which is in a state like (\ref{2a}):
\begin{equation}
\left| \Psi \right\rangle _{all}=(\alpha \left| in\right\rangle _{1}+\beta
\left| out\right\rangle _{1})\otimes (\alpha \left| in\right\rangle
_{2}+\beta \left| out\right\rangle _{2})\otimes ...\otimes (\alpha \left|
in\right\rangle _{n}+\beta \left| out\right\rangle _{n}).  \label{3}
\end{equation}
Intuitively, due to the fact that for such a state, according to the above
discussion, each marble\emph{\ is} inside the box, one would like to be
entitled to conclude that \emph{all marbles} are inside the box.

Lewis [1997] as well as Clifton and Monton [1999] call attention to the
fact that
according to any one of the previous criteria this cannot be the case and,
consequently, the GRW theory violates the enumeration principle. Actually
Lewis states:

\begin{quotation}
... the proximity of $|\Psi \rangle _{all}$ to the eigenstate of all $n$
marbles being in the box is given by
\begin{equation}
|_{all}\langle \Psi |all\,in\rangle |^{2}=\left| \alpha \right| ^{2n}.
\label{4}
\end{equation}

Recall that $\left| \alpha \right| ^{2}$ is slightly less than 1. This means
that as $n$ becomes very large, $\left| \alpha \right| ^{2n}$ becomes small;
... Consequently, when $n$ is sufficiently large $|\Psi \rangle _{all}$ is
not close to the eigenstate of all $n$ marbles in the box ... Given this, we
certainly cannot claim that $|\Psi \rangle _{all}$ is a state in which all $%
n $ marbles are in the box\footnote{%
We stress that the above expression (\ref{4}) is precisely the probability
that Standard Quantum Mechanics (SQM)\ with the Wave Packet Reduction
Postulate (WPR) attributes to the outcome $n$ of a measurement aimed to
ascertain how many particles are in the box.} (Lewis [1997], p.318).
\end{quotation}

Clifton and Monton [1999] argue in a similar way making reference to a
generalization of PosR which they denote as the \textit{fuzzy link}:

\begin{quotation}
`Particle $x$ lies in region $R_{x}$ \textit{and y} lies in $R_{y}$ \textit{%
and z} lies in $R_{z}$ and ...' if and only if the proportion of the total
squared amplitude of $\psi (t,\mathbf{r}_{1},\mathbf{r}_{2},...,\mathbf{r}%
_{n})$ that is associated with points in $R_{x}\times R_{y}\times
R_{z}\times ...$ is greater than or equal to $1-p$ (p.4)
\end{quotation}
and they conclude:

\begin{quotation}
...by applying the fuzzy link for $\left| \alpha \right| ^{2n}\leq p$ to $%
|\Psi \rangle _{all},$ one obtains the result that not all the marbles are
in the box. And this seems to contradict the results one obtains when one
applies the fuzzy link on a marble-by-marble basis, where one gets the
results that marble 1 is in the box, marble 2 is in the box, and so on
through to marble $n$ (p.9).
\end{quotation}

We stress that Lewis [1997] in his paper insists in adopting the scalar product
criterion to characterize the nearness of a statevector to another one and
makes reference to this formal `nearness' to attribute precise properties
concerning the location of a macroscopic physical system\footnote{%
To be precise, we have to mention that Lewis [1997] in his paper has taken
into account
the `objective mass density' interpretation of the GRW theory which will
represent the
basis of the analysis of the next section and the associated criterion of
`macroscopic
nearness' which has been introduced by Ghirardi, Grassi and Benatti [1995].
However, to
stress that these interpretations do not allow to overcome the
difficulties, he has to
resort to analyzing `the chance of finding all \textit{n} marbles on
observation'. This
position correponds to mixing up the ontology which is the appropriate one
for the GRW
theory  with the one which characterizes SQM with the WPR postulate, a
misleading
and inappropriate procedure.}. Similarly, Clifton and Monton in their paper
stick always to the \textit{fuzzy link }criterion. Instancies of this
precise attitude can be found, e.g., after their eq. (15) and in various
other places of their paper.

Before coming to analyze our answer to Lewis [1997], the criticisms to it by
Clifton and Monton [1999], and their central argument  aiming to prove that
the counting
anomaly cannot, in principle, ever become manifest, we consider it
important to call
attention on the peculiar attitude that the positions of Lewis [1997] and
of Clifton and
Monton [1999] imply with respect to the ontology of dynamical reduction models.

\section{The Interpretation of the GRW Theory}

In our opinion the most serious drawback of (Lewis [1997]) and (Clifton and
Monton
[1999]) derives from their misleading and inconsistent use of the dynamical
reduction
formalism. Such a formalism finds its very conceptual motivations in the
desire to
overcome the measurement problem of quantum mechanics. The most
characteristic trait of
the GRW theory consists, in J.S. Bell's words, in the fact that:

\begin{quotation}
There is nothing in this theory but the wavefunction. It is in the
wavefunction that we must find an image of the physical world, and in
particular of the arrangement of things in ordinary three-dimensional space
... [Schr\"{o}dinger] would have liked the complete absence of particles
from the theory, and yet the emergence of `particle tracks', and more
generally of the `particularity' of the world, on the macroscopic level
(Bell [1987],
p. 44).
\end{quotation}

Accordingly, there is no `measurement process' as distinct from any other
physical process. This fact is so universally recognized, even by those who
do not share the dynamical reduction point of view, that the GRW and CSL
theories have been classified among the `quantum theories without
observers'. Actually, as a consequence of a lively debate about the meaning
of the theory, the appropriate interpretation (now universally accepted) of
the theory has been precisely formulated by Ghirardi, Grassi and Benatti
[1995]. It is
based in an essential way on the consideration of the average mass density
c-number function $\mathcal{M}(\mathbf{r},t)$ in ordinary three-dimensional
space defined as:
\begin{equation}
\mathcal{M}(\mathbf{r},t)=\langle \Psi (t)|M(\mathbf{r})|\Psi (t)\rangle ,
\label{6}
\end{equation}
where $|\Psi (t)\rangle $ is the statevector describing the
individual physical system at time $t$ and $M(\mathbf{r})$ is the average
mass density operator at point \textbf{r}, the average being taken on the
typical localization volume $\left[ 1/\sqrt{\alpha }\right] ^{3}$
which characterizes the theory:
\begin{eqnarray}
M(\mathbf{r}) & = & \sum\nolimits_{k}m_{k}N^{(k)}(\mathbf{r}) \nonumber \\
& \equiv &
\sum\nolimits_{k}m_{k}\left\{ \left[ \frac{\alpha }{2\pi }\right] ^{\frac{3}{%
2}}\sum\nolimits_{s}\int d\mathbf{q}e^{-\frac{\alpha }{2}(\mathbf{q}-\mathbf{%
r})^{2}}a_{k}^{\dagger }(\mathbf{q},s)a_{k}(\mathbf{q},s)\right\} .
\label{7}
\end{eqnarray}
In the above equation $a_{k}^{\dagger }(\mathbf{q},s)$ and $a_{k}(\mathbf{q}%
,s)$ are the creation and annihilation operators of a particle of type
\textit{k} (\textit{k}  =  electron, proton,...) at point\textbf{\ q}, with
spin
component \textit{s}.

The most important feature of the GRW\ dynamics (contrary to what happens
within SQM) is that of making, in the case of macroscopic objects, the value
of the quantity $\mathcal{M}(\mathbf{r},t)$ \textit{objective} or, resorting
to a physically more expressive term which has been introduced by Ghirardi
and Grassi
[1996], \textit{accessible}. The very idea of accessibility is
appropriately expressed as follows:

\begin{quotation}
A property corresponding to a value (a range of values) of a certain
variable in a given theory is objectively possessed or accessible when,
according to the predictions of that theory, experiments (or physical
processes) yielding reliable information about the variable would, if
performed (or taking place), give an outcome corresponding to the claimed
value. Thus, the crucial feature characterizing accessibility (as far as
statements about individual physical systems are concerned) is the matching
of the claims and the outcomes of physical processes testing the
claims\footnote{We
stress the crucial role played in the above sentence by the specification
`according to
the prediction of that theory'. Obviously, we perfectly agree that any
theory has to be
subjected to experimental tests and, consequently, that those physical
processes designed
to test it have an extreme relevance. But we also stress that when a theory
allows to
make precise statements about physical properties (as it happens e.g., for
classical
mechanics) one does not need to actually perform a test every time one
makes claims
about the values  possessed by the physical quantities one is interested
in.} (Ghirardi
[1997], p. 227).
\end{quotation}

The fundamental feature of the GRW's dynamics of making accessible, in the
macroscopic case, precisely the mass density function $\mathcal{M}(\mathbf{r}%
,t)$ has been discussed in all details by Ghirardi, Grassi and Benatti
[1995]. In that
paper it has been proven that within such a theory any macroscopic body
ends up,
in extremely short times, in such a state that all physically testable
effects related to the mass density distribution actually agree with the
statement that such a distribution is the one which is actually present,
independently of any test being actually performed or not. This is why it is
generally accepted that the ontology of the dynamical reduction program
implies (at the macroscopic level) to answer to John Bell's [1990] question:
{\it Probability of what exactly are the probabilities of your theory?} by
stating:
probabilities that the mass density at the various points of ordinary space
be the one
given by
$\mathcal{M}(\mathbf{r},t)$. Ghirardi, Grassi and Benatti [1995] have
presented a precise
mathematical criterion to evaluate whether the mass density is accessible at
point\textbf{\ r}: one takes into account the ratio
$\mathcal{R}^{2}(\mathbf{r},t)$ of the variance
$\mathcal{V}(\mathbf{r},t)$ to the square of $\mathcal{M}(\mathbf{r},t):$%
\begin{equation}
\mathcal{R}^{2}(\mathbf{r},t)=\frac{\mathcal{V}(\mathbf{r},t)}{\mathcal{M}^{2}(\
mathbf{r},t)}%
\equiv \frac{\langle \Psi (t)|\left[ M(\mathbf{r})-\langle \Psi (t)|M(%
\mathbf{r})|\Psi (t)\rangle \right] ^{2}|\Psi (t)\rangle }{\langle \Psi
(t)|M(\mathbf{r})|\Psi (t)\rangle ^{2}}.  \label{8}
\end{equation}
Then, when $\mathcal{R}^{2}(\mathbf{r},t)<<<1$ the mass density at\textbf{\ r%
} is accessible. If the mass density is accessible at all points of a region
\textit{R}, all physical effects agree with the assumption that the
considered mass density objectively characterizes the situation within the
region.

The reader will easily understand (see  Ghirardi, Grassi and Benatti [1995]
for the
rigorous detailed proofs) that, for a state like (\ref{2a}), the mass
density is
accessible at all points of the volume of 1 cubic centimeter around the
centre of mass position (which, as we will show in section 4 turns out to be
extremely well definite) of the marble. Exactly the same argument shows that
at all space points lying outside this region the mass is non accessible.
Actually, by the argument of Ghirardi, Grassi and Benatti [1995] it can be
shown that
for all points `occupied' by the marble, $\mathcal{R}(\mathbf{r},t)\cong
e^{-\left[
10^{15}\right] }$ and that the total mass outside the just mentioned region
turns out to be of the order of $e^{-\left[ 10^{15}\right] }$ nucleon masses.

If one takes into account the correct perspective about the interpretation
of the GRW theory and the explicit proofs of all above mentioned facts, one
should have clear why the enumeration anomaly cannot arise within such a
theory. The very universal dynamics characterizing the theory guarantees
that in the state (\ref{3}) the mass is objective precisely in the regions
where the various marbles are, and that all physical tests one can imagine
will confirm precisely that the actual mass density distribution is the one
corresponding to the above statement. The marbles can therefore be claimed
to be all within the box, and the total mass within the box is actually the
one corresponding to all of them being in the box. There is absolutely no
need of actually performing any test to be sure that these statements are
correct: the very formal structure of the theory guarantees that this is the
case.

Obviously, due to the dynamics of the theory, even a state like (\ref
{3}) can change and, at a subsequent time (see however section 4 for a
detailed analysis) could evolve in a state of the type:
\begin{equation}
\left| \widetilde{\Psi }\right\rangle _{all}=(\alpha \left| out\right\rangle
_{1}+\beta \left| in\right\rangle _{1})\otimes (\alpha \left|
in\right\rangle _{2}+\beta \left| out\right\rangle _{2})\otimes ...\otimes
(\alpha \left| in\right\rangle _{n}+\beta \left| out\right\rangle _{n}).
\label{9}
\end{equation}
However, this fact (i.e., the change of the space region where the mass
density distribution referring to marble 1 is accessible) does not give rise
to any difficulty: for such a state the mass is objective where the marbles
are, i.e. around the centre of mass of the $n-1$ marbles in the box, and
around the centre of mass of marble 1 which is outside the box. And any
conceivable test, be it actually performed or not, will agree with these
statements.

Here a short digression to reply to one of the criticisms of Clifton and
Monton [1999] to our paper (Ghirardi and Bassi [1999]) is at order. When,
to describe
the above possible evolution of the statevector (\ref{2a}), in section 3 of our
paper we have made explicit the result of taking the products of the terms
of the state $\left| \Psi \right\rangle _{all}$%
\begin{eqnarray}
|\Psi \rangle _{all}  & = & \alpha ^{n}|in\rangle _{1}\otimes |in\rangle
_{2}...|in\rangle _{n}+\alpha ^{n-1}\beta |out\rangle _{1}\otimes |in\rangle
_{2}...|in\rangle _{n}+ \nonumber \\
& & +...\beta ^{n}|out\rangle _{1}\otimes |out\rangle
_{2}...|out\rangle _{n},  \label{10}
\end{eqnarray}
and we have stated that:

\begin{quotation}
the precise GRW dynamics will lead in about one millionth of a second to the
suppression of the superposition and to the `spontaneous reduction' of the
state $\left| \Psi \right\rangle _{all}$ to one of its terms (Ghirardi and
Bassi, p. 708),
\end{quotation}
we wanted simply to call attention to the fact that what could happen was
just the transition from a state like (\ref{3}) to one like (\ref{9}) --- or
even a similar one in which $k$ particles are outside the box. Actually,
Clifton and Monton [1999] have pointed out (more appropriately) that under
the GRW dynamics only states like (4) and (9) - or similar ones in which
more marbles are `out of the box' - can occur. Thus, one could say that we
have been a little bit sloppy in using the above expression `one of the
terms (of Eq. (10))' to make reference to states like (4) or (9). However,
the point we wanted to make is that a rigorous interpretation of states of
this type requires the consideration of the mass density functional (as we
have made clear in this Section) and cannot be based on the {\it fuzzy
link} criterion or other inappropriate criteria. Therefore, the criticism
of Clifton and Monton [1999] to our previous paper misses the crucial
point of our argument just because it does not make reference to valid
criteria for interpreting the wavefunction within dynamical reduction
models.

Concluding, within theories like the GRW model no violation of the enumeration
principle can occur if one correctly looks at it in terms of the
accessibility of the
mass density distribution. But something more must be said with reference
to the paper
by Clifton and Monton [1999]. Even though, as we have just discussed, they have
misinterpreted our argument, they agree with us in rejecting Lewis' [1997]
criticisms
because, in their opinion, if one

\begin{quote}
operationalizes the process of counting marbles by explicitly modelling the
process itself in terms of collapsing the GRW wavefunction (Clifton and
Monton [1999],
p.19),
\end{quote}
the violation of the enumeration principle cannot be, even in principle, put
into evidence. For those who have followed our discussion about the ontology
of the theory it should be clear that the GRW model does not in any way
require to operationalise the process of counting the marbles, since the
theory makes precise (and testable if one wants to do so) statements about
the actual location of the marbles at any time. This shows that the
arguments of Clifton and Monton [1999] are superflous. But this is not the
whole story: in section 5 we will show that the treatment of these authors
is unprecise and misleading.

Before coming to such an analysis we would like to present a very simple
study of the actual dynamics of a system like the one under consideration to
have the opportunity of analyzing better the various situations which can
occur. This will allow us to stress how careful one must be if one wants to
consider (unphysical) limits like those based on the consideration of a
number of marbles as large as one wants. In order to avoid being
misunderstood we warn the reader that we have no objections about taking the
considered limit, but we want to show how one has to deal with it if one
wants to argue in a consistent way. Our discussion will show that while the
GRW theory does not, in any case, meet difficulties with the enumeration
principle, the attempt to operationalize the counting of the particles can
become a nonsensical task. This is another reason which makes the analysis
of Clifton and Monton [1999] quite inappropriate.

\section{A Simplified Realistic Example of the GRW Dynamics for a
Macroscopic Object}

We consider it extremely useful, to give a precise sense to the problem
under investigation, to present an explicit analysis of the dynamical
evolution of a single macroscopic system such as one of the marbles of the
previous sections. Suppose that at the initial time ($t=0$) its normalized
centre-of-mass wavefunction is\footnote{%
For simplicity, we treat the system as if the centre-of-mass motion takes
place in one dimension. This assumption does not change in any way the
conclusions we will draw.}:
\begin{equation}
\Phi (x,0)=\left[ \frac{a}{\pi }\right] ^{\frac{1}{4}}e^{-\frac{a}{2}x^{2}}
\label{a}
\end{equation}
We disregard the internal motion and we confine our attention to the
spontaneous localization process. We recall that for a macro-object the
localization frequency $\lambda $ is amplified according to the number N of
its constituent nucleons, which, for the case under consideration (N$\simeq
Avogadro^{\prime }s$ number) means that:
\begin{equation}
\lambda \cong N\lambda _{micro}\simeq 10^{7}\sec ^{-1}.  \label{b}
\end{equation}
We consider now the effect of a localization, taking place at $x_{0}$ at
time $t_{1}$ on the state $\Phi (x,0).$ We have:
\begin{equation}
\Phi (x,0)\Rightarrow \Phi _{x_{0}}(x,t_{1})=\left[ \frac{a+\alpha }{\pi }%
\right] ^{\frac{1}{4}}e^{-\frac{a+\alpha }{2}(x-\frac{\alpha }{a+\alpha }%
x_{0})^{2}}  \label{c}
\end{equation}
Taking into account the various localizations processes which take place in
the interval $(0,t),$ the wavefunction $\Phi _{x_{0}(t)}(x,t)$ at time t
will be:
\begin{equation}
\Phi _{x_{0}(t)}(x,t)=\left[ \frac{a+n(t)\alpha }{\pi }\right] ^{\frac{1}{4}%
}e^{-\frac{a+n(t)\alpha }{2}(x-x_{0}(t))^{2}}  \label{d}
\end{equation}
In the above equation $n(t)$ is the number of localization processes in the
considered time interval. Such a number is a Poisson process with average
value $\langle n(t)\rangle =\lambda t.$ Similarly $x_{0}(t)$ is a stochastic
process with zero mean value: $\langle x_{0}(t)\rangle =0.$ The variance of
the final Gaussian wave function and the rate at which it decreases with
time are given by:
\begin{equation}
\sigma _{Loc}^{2}(t)=\frac{1}{a+n(t)\alpha }\simeq\frac{1}{a+\alpha \lambda t}%
;\qquad \frac{d\sigma _{Loc}^{2}(t)}{dt}\simeq -\frac{\alpha \lambda
}{(a+\alpha
\lambda t)^{2}}.  \label{e}
\end{equation}

The above equations show that the wavefunction tends, for $t\rightarrow
+\infty ,$ to a Dirac's delta function. However, one has to take into
account that the free dynamics of the centre of mass implies, according to
Schr\"{o}dinger's equation, an increase of the spread of the wavefunction
(which has always a Gaussian shape) which is more rapid the narrower is the
wavefunction itself. The rate of increase in the case under consideration is
given by:
\begin{equation}
\frac{d\sigma _{Sch}^{2}(t)}{dt}=\frac{4\pi ^{2}\hbar ^{2}(a+\lambda \alpha
t)t}{m^{2}}.  \label{f}
\end{equation}
A regime condition is reached when the rate of decrease due to the
localizations equals the rate of increase due to the free spread, which
happens for a time $\widetilde{t}:$%
\begin{equation}
\widetilde{t}=\sqrt{\frac{m}{2\pi \hbar \lambda \alpha }}\simeq 10^{5}\sec ,
\label{g}
\end{equation}
the indicated value referring to the case of a macroscopic object of the
kind we are considering. If this value is replaced into the expression for
the width one gets for the standard deviation in position:
\begin{equation}
\Delta x=\left[ \frac{2\pi \hbar }{m\lambda \alpha }\right] ^{\frac{1}{4}%
}\simeq 10^{-11}cm.  \label{h}
\end{equation}
It is useful to note that this value matches almost exactly the one which
has been identified by Ghirardi, Rimini and Weber ([1986], Section 8)  by a
much more
complex and realistic analysis. It has also to be remembered that in the
same paper it
has been proved that the GRW dynamics leads (obviously in the case of a
macroscopic system) to a momentum spread such that the Heisemberg relations
are almost satisfied with the equality sign.

The conclusion of this first part has to be kept in mind to grasp the real
situation one is dealing with: according to the GRW theory a macro-object
will find itself, just as a consequence of the dynamics governing all
natural processes, in a state such that its centre of mass wavefunction has
a spread of the order of $10^{-11}cm$. The typical wavefunction of the
centre of mass of each of our marbles will therefore be:
\begin{equation}
\Phi (x)=\left[ \frac{10^{22}}{\pi }\right] ^{\frac{1}{4}}e^{-\frac{10^{22}}{%
2}(x-\overline{x})^{2}}  \label{i}
\end{equation}
for an appropriate $\overline{x}.$

Starting with such a state, let us suppose that it suffers repeated
localizations all at the same place $x_{0}$ at a macroscopic distance from $%
\overline{x}.$ As we have seen, the wavefunction will keep its Gaussian
structure whose equilibrium width does not change, while its position
(taking into account that $10^{22}>>\alpha =10^{10}$ ) is shifted according
to:
\begin{equation}
\overline{x}\Rightarrow \widetilde{x}\simeq\overline{x}+\frac{\alpha
\lambda t}{%
10^{22}}x_{0}.  \label{l}
\end{equation}
Therefore, to have a displacement of the order of $x_{0}$ of the centre of
mass of one of our marbles it takes a time of the order of 1 day and a
number of localizations (all of them occurring around $x_{0})$ of the order
of $10^{12}.$ Since the probability of a localization at a distance larger
than $x_{0}-\overline{x}$ equals \textit{erfc}$[10^{11}(x_{0}-\overline{x}%
)], $ one sees that for a distance $x_{0}-\overline{x}$ of the order of 10
cm, such a probability is of the order of $e^{-[10^{22}]}.$ Concluding the
probability that our marble be displaced of 10 cm in one day is much smaller
than $e^{-[10^{34}]}!$

We stress once more that we have not made this calculation to prove how
unphysical is to think that a marble is drawn out of the box by the
spontaneous localizations, but to point out that such a process, which can
certainly occur in principle for an unphysically large number of particles,
requires a remarkable time. This means that, in the conditions we are
envisaging, the state will keep the form (\ref{3}) for quite long times and
that, provided the number of marbles is incredibly large, it will change,
e.g., into the state
\begin{equation}
\left| \Psi \right\rangle _{all}=(\beta \left| in\right\rangle _{1}+\alpha
\left| out\right\rangle _{1})\otimes (\alpha \left| in\right\rangle
_{2}+\beta \left| out\right\rangle _{2})\otimes ...\otimes (\alpha \left|
in\right\rangle _{n}+\beta \left| out\right\rangle _{n}).  \label{m}
\end{equation}
smoothly in time. Thus, the marbles go out of the box quite slowly.

In the just considered case the situation turns out to be perfectly
reasonable. At the beginning, starting from a state like (4), the mass will
be objective
at the points (within the box) where the marbles are, and correspondingly
the total mass
contained within the box will objectively match the one of the \textit{n}
marbles which are in it. The situation will change with time and, after some
time, one (or more) marble will go out of the box. Once more the theory
guarantees that the mass is accessible precisely in the regions where the
marbles are
located and guarantees the accessibility of the total  mass within the box
and the
matching of these accessible values. There is no need to test
these claims: they have been proved to be logically implied by the
formalism. Obviously,
if one wants to do so, one can check, by resorting to appropriate tests
which are
governed by the precise laws of the theory, that things are just as we have
stated even though this procedure is conceptually irrelevant. In spite of
the fact that tests for the positions of all marbles and for the total mass
within the box could require a remarkable time, let us assume that one can
cope with this program within the indicated time interval.

However we have now to take into account other possibilities. Since we have
accepted the challenge of assuming that the number of marbles is as large as
one wants, we have to point out that a single localization of a marble can
occur at a place which is very far from its centre of mass position.
Equation (\ref{l}) shows that a {\it single} localization taking place at a
point
separated from the centre of mass of the marble by a distance of 10$^{14}cm$
(i.e. of about 1 light-day) will displace the marble about 1 $m$.
Obviously, a localization has a probability of about $e^{-[10^{50}]}$ of
occurring at such a distance but, since we are playing the peculiar game of
putting no limit to the number of marbles, one can very well claim that some
marble will certainly jump in and out at every localization. And this remark
(even though we do not like to play with science fiction arguments) gives
more strenght to our position than to the one of Clifton and Monton [1999].

According to the GRW theory with its ontology, the situation is perfectly
clear: \thinspace at any time a certain number of marbles are in the box and
a certain number are out of the box and the corresponding statements about
the mass within and outside the box match with those concerning the
locations of the individual marbles. The situation, however, changes every
millionth of a second. Thus, if one thinks that to tackle the conceptual
problem of the counting anomaly requires a line (which we will analize in
the next
section) of the type of the one of Clifton and Monton [1999] of
operationalizing the
counting process through measuring instruments, then he must first of
all assume that it is possible to perform the incredible task of testing
the positions
of all particles and the total mass within the box in much less than a
millionth of a
second.

The general picture should now be clear: for any state like (21) the assertions
concerning each particle being in or out of the box and those about the
number of
particles which are in the box agree perfectly. The agreement is clearly
implied and
guaranteed by the very formal structure of the theory and by the only
physically
meaningful interpretation which allows to speak of accessible mass
distribution within the theory (i.e. by the very ontology of the model). To
reach this conclusion there is no need to actually consider practical ways
to perform tests identifying whether each given marble is in or out and the
total number of particles which are in the box. At any rate, if one has the
time and the opportunity to perform the tests, one will never meet a
contradiction with the enumeration principle.

We pass now to analyze the arguments of Section 4 of Clifton and Monton's
paper.

\section{ A
Critical Analysis of the Request to Operationalize the Process of Counting
Marbles}

The line chosen by Clifton and Monton [1999] to tackle the problem
under discussion consists in proving, by operationalizing the counting
process, that the violation of the enumeration principle within theories of
the GRW type cannot be put into evidence. In fact, as already mentioned,
the authors claim that

\begin{quotation}
The trouble is that Lewis fails to operationalize the process of counting
marbles by explicitly modelling \textit{the process itself} in terms of
collapsing GRW wavefunctions (Clifton and Monton [1999], p. 19).
\end{quotation}
The unappropriateness of such a point of view has already been stressed in
section 3. Within GRW's theory no \textit{measurement} process ever occurs,
the only physical processes being interactions among physical systems
governed by universal laws. In a case like the one under consideration,
since both the marbles to be counted and the measurement instruments which
are devised to count them are macroscopic systems which have to be put
exactly on the same grounds from the point of view of the spontaneous
localization processes affecting them, to operationalize the counting
process amounts to make uselessly more complicated the problem. The theory is
absolutely explicit about the position properties of the marbles and there
is no need
to resort to measuring apparata.

Leaving aside this obvious remarks, we feel the necessity of stressing
further serious drawbacks which affect the central section 4 of the Clifton
and Monton [1999] paper. In order to do this we have to follow the authors
in their attempt to operationalize the counting process. They consider
first of all%
\textit{\ n} apparatuses, one for each marble, devised to detect whether each
marble is or is not in the box. The process involves correlating the $%
|in\rangle $ and $|out\rangle $ states of the marble to orthogonal states of
a macroscopic measuring apparatus. This process should lead from the state
\begin{eqnarray}
\lbrack (\alpha \left| in\right\rangle _{1}+\beta \left| out\right\rangle
_{1})\otimes (\alpha \left| in\right\rangle _{2}+\beta \left|
out\right\rangle _{2})\otimes ...\nonumber \\
\otimes (\alpha \left| in\right\rangle
_{n}+\beta \left| out\right\rangle _{n})]  \label{n}
\otimes |ready\rangle _{M1}\otimes ...\otimes |ready\rangle _{Mn}
\end{eqnarray}
to the state
\begin{eqnarray}
(\alpha \left| in\right\rangle _{1}|IN\rangle _{M1}+\beta \left|
out\right\rangle _{1}|OUT\rangle _{M1})\otimes ... \nonumber \\
\otimes (\alpha \left|
in\right\rangle _{n}|IN\rangle _{Mn}+\beta \left| out\right\rangle
_{n}|OUT\rangle _{Mn})  \label{o}
\end{eqnarray}
with obvious meaning of the symbols.

Here we cannot avoid stressing a fundamental fact: the apparatuses, to allow
one to know whether they indicate IN or OUT must exhibit some macroscopic
difference\footnote{%
In fact, if the orthogonality of the final apparatus states would be
correlated, e.g., to one of their atoms being in the fundamental or in the
first excited state, respectively, such systems would not be of any use
for our purposes.}. Such macroscopic differences, in perfect agreement with
the GRW ontology, will be related to different locations of some
macroscopic part of the apparatuses themselves in order to allow the
observer's reading of the `pointer' to get the desired information.
However, since no state of a macroscopic object can have compact support
in configuration space, also the states of the apparatuses will be
superpositions of states corresponding to different locations of `their
pointer'. Thus, one must accept that the correct final apparatus states
will be of the type:
\begin{equation}
|IN\rangle_{Mi} \rightarrow |\widetilde{IN}\rangle _{Mi}=\gamma |IN\rangle
_{Mi}+\delta
|OUT\rangle _{Mi,\qquad }
\end{equation}
\begin{equation}
|OUT\rangle_{Mi} \rightarrow |\widetilde{OUT}\rangle _{Mi}=\delta
|IN\rangle _{Mi}+\gamma
|OUT\rangle _{Mi,\qquad }  \label{p}
\end{equation}
with $\left| \gamma \right| ^{2}>>>$ $\left| \delta \right| ^{2}.$ Thus,
what is the
usefulness of resorting to measuring instruments to identify the positions
of the
marbles if one has then to identify the position of the pointer
of the instrument  to become aware of the outcome?

According to Clifton and Monton [1999], once one has a state like
(\ref{o}), two
further steps are still necessary to acquire knowledge about the number of
marbles which are in the box. First of all one needs a further apparatus $M$
whose associated observable \textit{O} has $n+1$ eigenvalues $o_{i}$ where

\begin{quotation}
the $o_{i}$-eigenspace of the operator associated with $O$ is the subspace
spanned by all the terms in the superposition (16) [our eq. (\ref{o})] which
have as coefficient $\alpha ^{i}\beta ^{n-i}$ (Clifton and Monton [1999],
p. 21).
\end{quotation}
When this is done one gets a state which is a superposition of products of
the (marbles+counting apparata) states and states of the kind $|^{\prime
}O=k^{\prime }\rangle _{M}$ specifying that precisely \textit{k} marbles are
in the state $|in\rangle $ and the \textit{k} associated operators $M_{i}$
are in the state $|IN\rangle _{Mi}$ (obviously, if $k<n$ the state
associated to $|^{\prime }O=k^{\prime }\rangle _{M}$ will be a superposition
of states in which the individual marbles are in different positions but
precisely $k$ of them are in the state $|in\rangle ).$ Since the state is a
superposition, a violation of the enumeration principle is still there.
But, the authors state:

\begin{quotation}
this does not mean that a failure of the rules of counting has now become
manifest. (Clifton and Monton [1999], p. 21)
\end{quotation}

Why it is so? According to Clifton and Monton:

\begin{quotation}
[because] the [resulting] state is highly unstable given the GRW dynamics,
since we see from [its form] that it is an {\it entangled} superposition of
states
of macroscopic systems, where [its] various terms markedly differ as to the
location of the pointer on $M^{\prime }$s dial that register the value of $O
$ (Clifton and Monton [1999], p. 21)
\end{quotation}

Accordingly, the GRW dynamics dictates a collapse on one of the terms of the
superposition. If the collapse is on the term containing the state $%
|^{\prime }O=n^{\prime }\rangle _{M},$ then there is no problem: in it
exactly $n$ marbles are in the state $|in\rangle ,$ exactly $n$ apparata are
in the state $|IN\rangle ,$ and $M$ has registered $O=n.$ However, if a
different value of $O$ occurs, e.g., $O=k,$ reduction will produce an
entangled state. Once more this does not give rise to any problem :

\begin{quotation}
since its terms (pairwise) differ as to location of at least one of the
marbles, and since the $M_{i}$ apparatuses and marbles are macroscopic, ...,
there will a further quick, effective collapse to one of the terms [with
precisely\textit{\ k} marbles in state $|in\rangle $ and precisely\textit{\
k }$M_{i}$ apparatuses in state $|IN\rangle ]$ (Clifton and Monton [1999],
p. 22).
\end{quotation}
Thus, once more, the number of particles which are in the box matches the
reading of the apparatus $M.$ Concluding, the violation of the counting rule
is still there, but it cannot be revealed.

We stress that the above argument is quite puzzling: the reasons for which
the time hierarchy suggested by the authors should be respected are totally
obscure. The marbles are macroscopic systems and their masses may very well
be comparable or even larger that those of the pointers both of the
apparatuses $M_{i}$ and of $M$. Why the GRW dynamics has to be suspended up
to the time in which the pointer of $M$ is localized, and subsequently (in
the case that $O\neq n)$ up to the moment in which all pointers of the $%
M_{i} $'s are localized, is really a big mistery.

A final remark is at order. Since $M$'s reading is given by its pointer
location (as explicitly stated by Clifton and Monton [1999]), a consistent
use of
the GRW dynamics implies that \textit{it cannot} be perfectly localized: it
will be affected by the `tails problem' just as all macroscopic systems.
Thus, if one follows the authors in their strange argument, one should
conclude that (with a small probability) the state can be the one in which $n
$ marbles are in the state $|in\rangle ,$ $n$ apparatuses $M_{i}$ register $%
|IN\rangle ,$ but the pointer of the apparatus $M$ can be reduced on a state
corresponding to its pointing at $k\neq n!$

The reader should have understood that the reasons for the peculiar account
of the whole process described by Clifton and Monton [1999] derive uniquely
from
the fact that the authors have betrayed, from the very beginning, the real
spirit of the dynamical reduction models, i.e. the basic fact that what
these models claim to be true of the world out there is the mass
distribution in the whole universe in those regions in which it is
accessible. All arguments must be developed keeping in mind this fundamental
point which Lewis [1997] and Clifton and Monton [1999] have disregarded. If one
does so, a very clear and simple picture of the problem emerges, and no
conceptual difficulties whatsoever arise.

\begin{center}
 \textbf{Acknowledgements}
 \end{center}
 We acknowledge useful discussions with Dr. Silvia Pascoli.\vspace{.3in}

 \noindent \emph{Department of Theoretical Physics,
 University of
Trieste, Italy \\(e-mail:
bassi@ts.infn.it)}.\vspace{.3in}

\noindent\emph{Department of Theoretical Physics,
 University of
Trieste, and \\ International Centre for Theoretical Physics, Trieste,
Italy \\(e-mail:
ghirardi@ts.infn.it)}.\vspace{.3in}

\begin{center}
 \textbf{References}
 \end{center}

\noindent Albert, D. and Loewer, B. [1996]:
`Tails of Schr\"{o}dinger's Cat',  in R.
Clifton (\emph{ed.}), \emph{Perspectives on Quantum Reality}, Dordrecht:
Kluwer,
pp. 81-91.\vspace{.1in}

\noindent Bell, J. S.  [1987]: `Are there Quantum Jumps?', in C. W.
Kilmister {\it (ed.)},
{\it Schr\"{o}dinger: Centenary Celebration of a Polymath}, Cambridge:
Cambridge
University Press, pp. 41--52. \vspace{.1in}

\noindent Bell, J. S.  [1990]: `Against Measurement', in A. I. Miller {\it
(ed.)}, {\it
Sixty--Two Years of Uncertainty}, New York: Plenum Press, pp. 17--31.
\vspace{.1in}

\noindent Clifton, R. and Monton, B. [1999]: `Losing Your Marbles in
Wavefunction Collapse
Theories', \emph{The British Journal for Philosophy of Science}, to appear.
Preprint quant--ph/9905065. \vspace{.1in}

\noindent Ghirardi, G.C. [1997]: `Macroscopic Reality and the Dynamical
Reduction
Program', in M.L. Dalla Chiara et al. {\it (eds.)}, \emph{Structures and
Norms in
Science}, Dordrecht: Kluwer,  pp. 221-240.  \vspace{.1in}

\noindent Ghirardi, G. C. and  Bassi, A. [1999]:
`Do Dynamical Reduction Models Imply That Arithmetic Does Not Apply to Ordinary
Macroscopic Objects?',
\emph{The British Journal for Philosophy of Science}, \textbf{50}, pp.
705-20.
\vspace{.1in}

\noindent Ghirardi, G. C. and Grassi, R. [1996]: `Bohm's Theory versus
Dynamical
Reduction', in J. T. Cushing {\it et al. (eds)}, \emph{Bohmian Mechanics
and Quantum
Theory: an Apprisal}, Dordrecht: Kluwer, pp. 353--377. \vspace{.1in}

\noindent Ghirardi, G. C., Grassi, R. and Benatti, F.  [1995]: `Decsribing the
Macroscopic World: Closing the Circle within the Dynamical Reduction
Program', {\it
Foundations of Physics} {\bf 25}, pp. 5--38. \vspace{.1in}

\noindent Ghirardi, G.C., Pearle, P. and Rimini, A. [1990]: `Markov
Processes in Hilbert
Space and Continuous Spontaneous Localization of Systems of Identical
Particles',
\emph{Physical Review A}, {\bf 42}, pp. 78--89. \vspace{.1in}

\noindent Ghirardi, G. C., Rimini, A., and Weber, T.  [1986]: `Unified
Dynamics
for Microscopic
and Macroscopic Systems', \emph{Physical Review D}, \textbf{34},
pp. 470-91.\vspace{.1in}

\noindent Lewis, P.  [1997]:
`Quantum Mechanics, Orthogonality, and Counting',
\emph{The British Journal for Philosophy of Science}, \textbf{48},
pp. 313-28.\vspace{.1in}

\noindent Pearle, P. [1989]: `Combining Stochastic Dynamical State--Vector
Reduction with
Spontaneous Localizations' {\it Physical Review A}, {\bf 39}, pp. 2277--89.
\vspace{.1in}

\noindent Shimony, A. [1991]: `Desiderata for a Modified Quantum Dynamics',
in A. Fine,
M. Forbes and L. Wessel {\it (eds)},
\emph{PSA 1990, Vol. 2}, East Lansing, MI: Philosophy of Science Association,
pp. 49--59. \vspace{.1in}

\end{document}